\documentstyle[aps,pre,multicol,epsf]{revtex}
\def\be{\begin{equation}}
\def\ee{\end{equation}}
\def\one{\hbox{$\bigcirc$\hspace{-0.3cm}{\em 1}}\hspace{0.25cm}}
\def\two{\hbox{$\bigcirc$\hspace{-0.3cm}{\em 2}}\hspace{0.25cm}}
\def\three{\hbox{$\bigcirc$\hspace{-0.3cm}{\em 3}}\hspace{0.25cm}}

\begin{document}

\title{Isotropic Transverse XY Chain with Energy- and Magnetization Currents}
\vspace {1truecm}

\author{T. Antal$^1$, Z. R\'acz$^{1,2}$, A. R\'akos$^1$, and G. M. Sch\"utz$^3$}

\address{{}$^1$Institute for Theoretical Physics,
E\"otv\"os University, 1088 Budapest, Puskin u. 5-7, Hungary\\
{}$^2$Department of Theoretical Physics,
University of Oxford, 1 Keble Road, Oxford, OX1 3NP, United Kingdom\\
{}$^3$Institut f\"ur Festk\"orperforschung, 
Forschungszentrum J\"ulich, 52425 J\"ulich, Germany}

\date{\today}

\maketitle

\begin{abstract}
{The ground-state correlations are investigated for 
an isotropic transverse XY chain which is constrained to carry either a
current of magnetization $(J^M)$ or a current of energy $(J^E)$. 
We find that the effect of $J^M\not= 0$ on the 
large-distance decay of correlations is twofold: i) oscillations 
are introduced and ii) the amplitude of the power law decay increases 
with increasing current. The effect of energy current is more complex.  
Generically, correlations in current carrying states are found to decay
faster than in the $J^E = 0$ states, contrary to expectations that correlations are increased by the presence of currents. However,
increasing the current, one reaches a special line where the correlations 
become comparable to those of the $J^E=0$ states. On this line, the 
symmetry of the 
ground state is enhanced and the transverse 
magnetization vanishes. Further increase of the current 
destroys the extra symmetry but the transverse magnetization remains at
the high-symmetry, zero value.  
}~
\end{abstract}
\begin{multicols}{2}
\narrowtext
\section{Introduction}
\label{sec:intro}
A general feature of nonequilibrium steady states is the presence of 
currents (fluxes) of some physical quantities such as energy, momentum, 
charge, etc.
Thus the study of nonequilibrium steady states is, in some sense, a study 
of the effects of currents imposed on the system either by  
boundary conditions and driving fields or by competing dynamical 
processes. An interesting and much investigated effect of
currents is the rather dramatic 
change in correlations. Namely, short-range 
correlations appear to change into long-range, 
power-law ones as the currents are switched on \cite{{liquid},{Zia}}. 
This is not entirely 
surprising in case of a global current since some conserved 
quantity is carried
fast compared to diffusion and, in the absence of detailed balance, 
this may generate long-range effective 
interactions and, as a consequence, long-range
correlations may appear \cite{SBR}.  
It is clear, however, that the general picture 
should be more complicated since 
large currents often lead to chaotic behavior 
which in turn results in weakened correlations \cite{Benard}. 
Thus, we believe, the understanding of the interplay of currents 
and correlations in nonequilibrium steady states
is a rather interesting and important task.

In order to achieve progress one tries to investigate simple models
and, indeed, a large number of classical statistical models have 
been introduced
for the study of nonequilibrium steady states \cite{Priv}. Unfortunately, 
far from equilibrium, classical systems are not constrained by
conditions like detailed balance and there is 
much arbitrariness in defining the dynamics. In order to avoid such
arbitrariness,
we have started to investigate quantum systems \cite{trising} 
where the time evolution is defined 
without ambiguity by the usual rules of quantum mechanics. 

Nonequilibrium steady states in a quantum
system may be investigated by imposing a current on the system and studying 
the properties of the ground state thus generated. 
As an example, we studied the
transverse Ising model \cite{trising} and
found that, in the presence of an energy current, the exponentially 
decaying two-spin correlations changed into power-law form thus supporting 
the notion that switching on currents increases correlations. 
In view of the lack of
detailed knowledge of the interplay of currents and correlations, 
here we probe the
generality of the above result on the example of 
the isotropic transverse XY chain. 
In this system, we have a global conservation not only for 
the energy - as in case of
the transverse Ising model - but also for the 
transverse magnetization and, consequently, 
the effect of both the energy- and magnetization currents can be investigated.
The XY chain is also interesting because it has 
power-law correlations already in the equilibrium state 
(i.e. in the state without any current) and so one might expect 
the system to be more sensitive to the introduction of currents.  
Indeed, in case of the energy current we find a rather complex behavior 
(including an increase of the ground-state symmetry at 
special values of the current) 
and our findings for the 
steady state correlations are at variance with those for the 
transverse Ising model. On the other hand, the changes we observe 
in the correlations 
due to a magnetization current allow for a straightforward 
interpretation in terms of 
increasing correlations due to presence of a current.

The effects of magnetization current are treated exactly (Sec.II) 
while the correlations in the presence of an energy current are
calculated by a combination of analytical and exact numerical methods in Sec.III.
Summary and final comments are contained in Sec.IV.

\section{Magnetization Current}
\label{sec:mag}

Our starting point is the Hamiltonian of the
$d=1$ isotropic XY model in a transverse field $h$:  
 \begin{equation}
 H^{XY}=-\sum_{\ell=1}^N \left( s^x_\ell s^x_{\ell+1}
 + s^y_\ell s^y_{\ell+1}  + h s^z_\ell \right)
 \label{trxyhami}
 \end{equation} 
where the spins are represented by Pauli spin matrices
$ s^\alpha_\ell$ ($\alpha =x,y,z$) at sites
$\ell=1,2,...,N$ of a one-dimensional periodic chain
($s^\alpha_{N+1}= s^\alpha_1$). The
transverse field, $h$, is measured in units of the Ising coupling, $J$, which 
is set to $J=1$ throughout this paper.
This model can be solved exactly \cite{LSM} since it
can be transformed by a Jordan-Wigner transformation into a set of free fermions with wavenumber $k$ and of energy 
 \be
 \Lambda_k = -\cos k - h.
 \label{spec}
 \ee

In this model, not only the total energy but also
the $z$ component of the total magnetization $M^z=\sum_\ell s_\ell^z$ 
is conserved.  
As a result, one can write down 
a continuity equation for the local magnetization $s_{\ell}^z$:
 \be
 \dot{s}_{\ell}^z = i[H^{XY}, s_{\ell}^z] = 
 j^M_{\ell} - j^M_{\ell-1}
 \ee
and this defines the magnetization current through the bonds:
 \begin{equation}
 j^M_{\ell} = s_\ell^y s_{\ell+1}^x  -  s_\ell^x s_{\ell+1}^y
 \label{current}
 \end{equation}
A macroscopic current can now be defined as  
\be
J^M=\sum_{\ell=1}^N j^M_\ell 
\ee
and one can recognize $J^M$ as the Hamiltonian of the  
Dzyaloshinskii-Moriya interaction \cite{DM}. Somewhat surprisingly,
the same expression 
emerged as the energy current in the case of 
the transverse Ising model \cite{trising}.

Our aim is to find the lowest energy state
among the states carrying a given current. 
Since $[H^{XY}, J^M]=0$, the problem can be solved using 
the Lagrange multiplier method,
i.e.\  we diagonalize the following Hamiltonian:
 \begin{equation}
 H^M = H^{XY} - \lambda J^M
 \label{newhami}
 \end{equation}
where $\lambda$ is a Lagrange multiplier. The ground state of $H^M$ can 
be considered as a current--carrying 
steady state of $H^{XY}$ at zero temperature.
Note that, without loss of generality, we can assume
$h \ge 0$ and $\lambda \ge 0$. 

The Hamiltonian $H^M$ is diagonalized using the same
transformations which diagonalize $H^{XY}$, 
and we get the following spectrum in the thermodynamic limit:
\be
\Lambda_k = \frac{1}{\cos \varphi} 
\left[ -\cos(k - \varphi) - \tilde{h} \right] ~,
\label{newspec}
\ee
where $\varphi=\arctan \lambda$ and an effective field 
$\tilde{h}=h\cos \varphi$ has been introduced. 
One can see that the spectrum is similar to that of $H^{XY}$ with the
wavenumber shifted by $\varphi$. It should be mentioned here that the above 
result in not new. It is implicit in the Bethe-ansatz solution of the 
anisotropic Heisenberg chain with twisted toroidal 
boundary conditions \cite{Alcaraz} and appears in various forms 
in studies of the Dzyaloshinskii-Moriya interaction \cite{DZstudies} and 
of the associated Berry phase \cite{Schutz1}.  

It is remarkable that $H^M$ can be transformed into $H^{XY}$ 
(\ref{trxyhami}) using a unitary transformation:
 \begin{equation}
 Q = e^{i\sum_{\ell=1}^N \ell \varphi s_{\ell}^z }
 \label{trans}
 \end{equation}
This transformation rotates the $\ell$th spin around the $z$ 
axis by angle $\ell \varphi$ and shifts the spectrum (\ref{newspec}) 
by the wavenumber $\varphi$. This shift is analogous to the phase shift experienced 
by electrons on a ring threaded by a constant magnetic flux 
$(\Phi \sim N\varphi)$ \cite{flux}. 

We also note that the ferro- and the anti-ferromagnetic cases
are equivalent in the sense that the 
canonical transformation (\ref{trans}) with $\varphi=\pi$ 
transforms them into each other. For a finite periodic chain, $H^M$
is transformed to $H^{XY}$, but with a twisted
boundary condition ($\varphi\not= \pi )$.

As the transformation (\ref{trans}) does not change the $z$ component of
the spins, we find that 
\be
\langle s_\ell^z \rangle = \frac{1}{\pi}\arcsin 
\frac{h}{\sqrt{1+\lambda^2}} \quad , 
\ee
and the correlation function
\be
\rho^z(r) = \langle s_\ell^z s_{\ell+r}^z\rangle=
-\frac{1}{\pi^2r^2}\sin^2\left[{r\arccos{\frac{h}{\sqrt{1+\lambda^2}}}}
\right ]
\ee
have their equilibrium form but at a different field
$\tilde{h}=h/\sqrt{1+\lambda^2}$ ($\langle ~~ \rangle$
denotes the expectation value in the ground state of $H^M$). 
One can see that, for $h\ge\sqrt{1+\lambda^2}$ ($\tilde{h}\ge1$),
the spins are parallel
to the field and there is no current in the system.
For  $h<\sqrt{1+\lambda^2}$ ($\tilde{h}<1$) the current has a simple form:
 \begin{equation}
 j^M = \langle J^M/N \rangle = { { \lambda \sqrt{1+\lambda^2-h^2}} \over  
                        {\pi (1+\lambda^2)} } ~,
 \end{equation} 
and one can observe that the maximum current, 
reached in the limit of $\lambda \rightarrow \infty$,
is given by $1/\pi$.

The introduction of the transformation (\ref{trans})
simplifies the calculation of the correlations
$\rho^x(r)=\langle s_\ell^x s_{\ell+r}^x\rangle = \rho^y(r)$ 
in the ground state of $H^M$:
 \begin{eqnarray}
 \rho^x(r) =  \cos(r \varphi) \langle Q \psi_0 
	| s_\ell^x s_{\ell+r}^x | Q \psi_0 \rangle \cr
 + \sin(r \varphi)  \langle Q \psi_0 
	| s_\ell^x s_{\ell+r}^y | Q \psi_0 \rangle 
  \label{crosscorr}
 \end{eqnarray}
where $\psi_0$ is the ground state of $H^{M}$, while $Q \psi_0$ is that 
of $H^{XY}$ at
a field $\tilde{h}$. Without any current, we have
$\langle  s_\ell^x s_{\ell+r}^y \rangle=0$ and, 
furthermore, the $r\rightarrow\infty$ 
behavior of the $\rho^x(r)$ correlation function
\cite{XYcorr} is given by 
 \be
 \rho^x\left(r; j^M\!=\!0 \right)\approx \left(1-h^2 \right)^{1/4} 
 {C \over \sqrt{r}} ~,
 \label{eqrhox}
 \ee
where $C=e^{1/2}2^{-4/3}A^{-6} \approx 0.147$ ($A \approx 1.282$ is the 
Glaisher's constant). 
Using (\ref{crosscorr}), one can obtain then the following simple form in 
the $r\rightarrow\infty$ limit 
 \be
 \rho^x(r) \approx \left(1-\tilde{h}^2 \right)^{1/4} 
 {C \over \sqrt{r}}
 ~\cos(r \varphi).
 \ee
Thus we find that the correlations decay by power law and they show 
oscillatory behavior in the current-carrying states. 
This is similar to what has been
observed in the transverse Ising model but there are some differences. 
In the Ising case, an exponential decay of correlations changes into a
power-law form as the current is switched on. 
In the XY case, on the other hand, one has power-law correlations already 
in the equilibrium state. The magnetization current does not change the 
power law and leaves its exponent intact as well. The increase of 
correlations appears as the increase in the amplitude of the power law
(note that $\tilde h = h\cos \varphi\le h$).

\section{Energy Current}

Since the energy is a conserved quantity as well, 
one can investigate the effects of the energy current.
The local energy (the contribution of the $\ell$th spin to the total energy) 
satisfies a continuity equation with the local energy current $j_\ell^E$,
and its sum over $\ell$ (the total energy current) has the form:
 \be
 J^E=\sum_{\ell=1}^N [s_\ell^z (s_{\ell-1}^y s_{\ell+1}^x -
 s_{\ell-1}^x s_{\ell+1}^y) + h( s_{\ell}^x s_{\ell+1}^y - 
 s_{\ell}^y s_{\ell+1}^x)]
 \label{jeop}
 \ee
One can easily show that $[H^{XY},J^E]=0$, and diagonalizing the 
Hamiltonian 
 \be
 H^E=H^{XY}-\lambda J^E ~,
 \ee  
one obtains the lowest energy eigenstates of $H^{XY}$ 
in the presence of a given $J^E$. 

Using the standard transformations to fermions 
again, the spectrum is obtained as:
 \be
 \Lambda_k = (-\cos k - h) (1-\lambda \sin k)~, 
 \label{spec2}
 \ee
and the modes with negative energy are occupied in the 
ground state of $H^E$. Although
the $k \rightarrow -k$ symmetry of the spectrum is broken for $\lambda \neq 0$, the ground state remains
that of $H^{XY}$ for $\lambda\le 1$ 
and, accordingly, no energy current flows through the 
system. This rigidity of the ground state against $\lambda$ is a consequence
of the fact that the fermionic spectrum of $H^E$ has a product form 
(\ref{spec2}), and the second factor is positive for $\lambda<1$. 
The first and second factors in $\Lambda_k$ change sign at 
$\pm(\pi/2+k_h)$ (for $h\le 1$)  
and $\pi/2 \pm k_\lambda$ (for $\lambda \ge 1$), respectively. 
The `critical momenta' $k_h$ and $k_\lambda$ are defined here 
such that
they take values $0 \leq k_h,k_\lambda \leq \pi/2$ and one has
$k_h = \arcsin{(h)}$ and $k_\lambda = \arccos{(\lambda^{-1})}$.

One can study the ground state as a function of $h$ and $\lambda$,
but we are more interested in the physical quantities as functions of
$h$ and $j^E = \langle J^E/N \rangle$.
Thus first we calculate $j^E$:
 \be
 j^E=\left\{ \begin{array}{ll} 
 \frac{1}{2\pi}(1+h^2-\lambda^{-2})~~~ & 
      \mbox{for $k_h \le k_\lambda$} \cr
&\cr 
 \frac{h}{\pi}\sqrt{1-\lambda^{-2}}  & 
      \mbox{for $k_h \ge k_\lambda$ or $h$, $\lambda \ge 1$}   \cr
&\cr
 0  & \mbox{for $\lambda\le 1$}, 
 \end{array} \right.
 \label{je}
 \ee
and than express all the $\lambda$ dependences in terms of $j^E$.
We can then obtain a $h-j^E$ phase diagram as shown on Fig.\ref{hjfig}
where the phases -- discussed below in more detail --
are distinguished by symmetries of the regions of occupied
states in the $k$-space.

\subsection{Phase diagram}

\begin{figure}
\epsfxsize=8truecm
        \epsffile{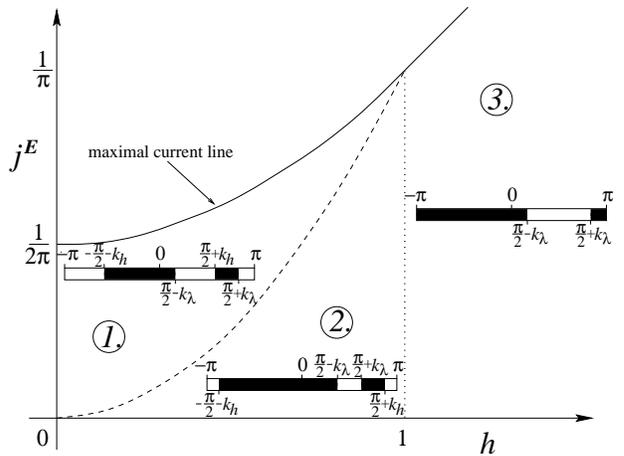}
\vspace{0.3cm}
\caption{\protect \narrowtext 
Phase diagram of the ground state of the transverse XY model
in presence of energy current. The black parts of the rectangles denote
the wavenumbers of the occupied fermionic modes ($-\pi\le k<\pi$).
The dashed line is a high-symmetry transition line between regions of \one
($M^z=0$) and \two ($M^z\ne 0$). There are no states above the 
maximal current line and region \three 
can be mapped onto the vertical dotted line ($h=1$).}
\label{hjfig}
\end{figure}

Let us begin the analysis of the phase diagram by first describing it 
in terms of the behavior of currents and of the transverse magnetization.
As shown in Fig.\ref{hjfig}, there is 
a maximal current for every value of $h$ 
 \be
 j^E_{max} = \left\{ \begin{array}{ll}
 (1+h^2)/(2\pi) & \mbox{for } h\le1 ~, \cr
 h/\pi & \mbox{for } h\ge 1 ~, \end{array} \right. 
 \ee
and no state exists above $j^E_{max}$.

We can divide the $h$--$j^E$ phase diagram into three regions.
The only interesting 
areas are \one and \two and their boundaries.
In region \three, the ground state is the same along the 
$j^E=\mbox{constant}\times h$ lines ($h\ge 1$) thus the $h=1$ line contains 
all the information about this region. Below, we restrict the 
discussion to the $h\le 1$ part of the phase diagram with the understanding 
that the $h=1$ line represents region \three .

As can be seen from (\ref{jeop}), the energy current has two parts: 
the term containing $h$ is proportional to the 
magnetization current ($-hJ^M $) while
the other term is the current of the interaction 
energy. 
The distinguishing features of regions \two and \three are 
that the current of interaction energy is zero  
\cite{magcurr} while the transverse magnetization, $M^z$,
is nonzero in the ground state. 

For any fixed value of $h$, the magnetization decreases 
with increasing $j^E$ and 
$M^z$ becomes zero on the line $k_h=k_\lambda$ corresponding 
to $j^E = h^2/\pi$. On this line, the magnetization-current part of $j^E$
saturates and, upon increasing $j^E$, we enter region \one where 
the interaction part of the current starts to flow. Another 
characteristic feature of region \one is that $M^z=0$ 
throughout this region. 

One tends to conclude at this point that the line separating 
regions \one and \two 
is a line of second 
order, nonequilibrium phase transitions with $M^z$ being the order parameter.
This notion is also supported by the facts that several quantities
such as $\rho^z(1)$ and $\rho^x(1)$ have a jump in their first derivatives 
when crossing the transition line and, 
furthermore, that the correlations are enhanced 
(change from $r^{-1}$ to $r^{-1/2}$) on this line (see Sec.\ref{sec:corr}). 
If this was a phase transition, however, it was certainly a transition not 
in the usual sense. The symmetry of the ground state
is the same on both sides of the transition line and 
the $M^z=0$ result in region 
\one is not a consequence of the up-down symmetry of the ground state.
The magnetization is zero in \one because the motion of the 
zeros of the dispersion relation (\ref{spec2}) conspire to keep the ground 
state at half-filling. We emphasize, however, that the half-filling 
does not mean that the ground state has a symmetry with respect 
to global spin-flip $s^z_i\rightarrow -s^z_i$.

It is interesting to note that the symmetry
of the ground state is higher on the transition line than on either side of it.
Indeed, on this `high'-symmetry line, the ground state is symmetric with 
respect to rotation of the spins around the $x$ axis by $\pi$,   
followed by a spatial reflection mapping site $i$ to $L+1-i$. 
The Hamiltonian, $H^{XY}$,
has no such symmetry and, off the transition line, the 
ground state doesn't have such symmetry either.  
Thus we can see here an example where the
increase of current in a system leads, at a particular value of the 
current, to symmetry enhancement. 
The reason for increase of symmetry is obviously some level crossing 
coming from the interplay of the current operator and the original 
Hamiltonian. One might speculate that the occurance of 
such symmetry enhancements is not an accidental but a general
feature of current carrying systems.

\subsection{Correlations}
\label{sec:corr}

The  $\rho^z(r)$ correlations can be calculated easily and, 
just as in the equilibrium case, one finds 
$\rho^z(r)\sim r^{-2}$. The difference from the equilibrium 
is that the oscillatory 
modulation of the $r^{-2}$ decay (present in equilibrium for $h\not= 0$)
becomes more complex. Such modulation has been observed 
in case of the imposed magnetic current (see Sec.\ref{sec:mag}) as well
as in the  transverse Ising model with energy current \cite{trising}.
The exponent of the power law decay, however, is unchanged when the 
currents are introduced in all of the above examples.
Thus it seems that $\rho^z(r)$ correlations are not too
sensitive to the presence of currents. 
A possible reason for this apparent rigidity is, perhaps, the lack of 
internal interactions among the $z$-components of the spins. 

It is harder to calculate the $\rho^x(r)=\rho^y(r)$ correlations 
but they show a more interesting behavior.
Some of our results described below are exact and were derived by combining
the Wick theorem and the spin rotation transformation (\ref{trans}) which 
relates the correlation functions between states where the 
ground-state occupation pattern in $k$-space is identical up to shifts 
$k\to (k+\alpha) \, mod\, 2\pi$. 
As we shall see, these exact results are restricted to the 
boundaries of regions \one and \two. At a general point $(h, j^E)$, we were 
able to calculate $\rho^x(r)\equiv \rho^x(r;h,j^E)$ 
numerically (for $r\le 100$ lattice spacings)
using the fact that the square of the 
correlation can be expressed as a determinant of a $2r\times 2r$ 
matrix with exactly calculable elements. 
 
Let us start by enumerating the exact results. 
The boundaries of region \two are discussed in points 1-3 while the boundaries 
of \one are treated in points 3-5.
\begin{enumerate}
\item
As discussed in Sec.\ref{sec:mag}, the correlations in the transverse 
XY model without current, i.e. on the 
line $(0<h<1, j^E=0)$, are known \cite{XYcorr} and the $r\rightarrow \infty$
asymptotics of $\rho^x$ is given by
 \be
\rho^x(r;h,j^E=0)\sim C\left(1-h^2 \right)^{1/4}r^{-1/2}~. 
 \label{jzerobound}
 \ee

\item
The correlations on the $h=1$ line can be related to the equilibrium 
case $(0<h<1, j^E=0)$ and one finds:
 \be
 \rho^x\left(r;1,j^E \right)=
 \rho^x\left(r;\sqrt{1-\pi^2{j^E}^2},0 \right)
 \cos \left(\frac{\pi}{2}r \right)~.
 \label{exacth=1}
 \ee
Thus, the large-distance behavior is given by 
\be
 \rho^x\left(r;1,j^E \right)= C\sqrt{\pi j^E}r^{-1/2}\cos(\frac{\pi}{2}r)~. 
\ee
Since the whole \three phase can be projected onto the $h=1$ line, 
we find that correlations decay as $r^{-1/2}$ for $h\ge 1$.
\item
The correlations on the `high-symmetry' line 
(i.e on the boundary between \one and \two) 
can also be related to equilibrium:
 \begin{eqnarray}
\rho^x(r;h,j^E=h^2/\pi)&=\rho^x(r;0,0) \cos(k_h r) \cr
&\cr
                       &\sim Cr^{-1/2}\cos(k_h r)
\label{bound}
 \end{eqnarray}
so we find again a $r^{-1/2}$ decay in 
the $r\rightarrow \infty$ limit.

\item
On the line of maximal-current $[h\le 1,j^E_{max}=(1+h^2)/(2\pi)]$, 
the correlations can be expressed 
in terms of those on the line $(h=0, j^E)$: 
 \be
 \rho^x(r; h,j^E_{max})=
 \rho^x\left(r; 0,\frac{1-h^2}{2\pi} \right) 
 \cos \left( \frac{\pi}{2}r \right) ~.
 \ee 
Unfortunately, this does not help in calculating the 
$r\rightarrow \infty$ behavior.

\item
The long range behavior of the correlations at the intersection 
point $[0,1/(2\pi)]$ of the 
$(h=0, j^E)$ and the $(h,j^E_{max})$ lines is also calculable:
 \be
 \rho^x\left(r; 0,\frac{1}{2\pi} \right)=
 \left\{ \begin{array}{ll} 4\left[ \rho^x(\frac{r}{2};0,0) \right]^2\sim
\frac{1}{r}~
 			      & \mbox{$\frac{r}{4}$ integer} \cr
                           0  & \mbox{otherwise}
	 \end{array} \right.
 \ee
and it is remarkable that the correlation function in this 
point decays as $1/r$ instead of $1/\sqrt{r}$. 
 
\end{enumerate}

The exact results can be summarized as follows. The $\rho^x=\rho^y$ 
correlations decay as $1/\sqrt{r}$ on the boundaries of 
region \two while a $1/r$ decay can be observed
in the upper-left corner $[0,1/(2\pi)]$ of the phase diagram.

\begin{figure}[htb]
\epsfxsize=8truecm
\epsfbox{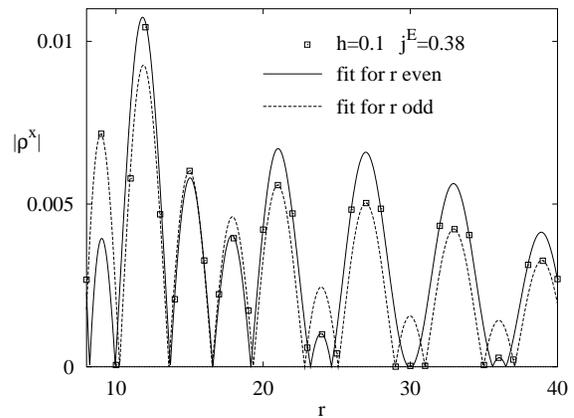}
\vspace{0.3cm}
\caption{\protect \narrowtext Absolute value of the 
$\rho^x(r)=\langle s^x_\ell s^x_{\ell+r} \rangle$ correlation function 
at a general point of the phase diagram. The distance $r$ is measured 
in units of lattice spacing. The long-range decay 
of $\rho^x(r)$ is fitted by the expression (\ref{bela}) (solid lines)
(for this point of the phase diagram one has 
$a_1 \approx a_2 \approx 0.0934$, $a_3 \approx 0.0469$). 
}
\label{fitfig}
\end{figure}

Numerical calculations suggest, however, 
that the $1/r$ asymptotics is more general than it looks 
from the exact results.
There is a strong indication that the large-distance asymptotics is actually
$1/r$ everywhere in region \one and \two 
apart from the boundaries of region \two.
Fig.\ref{fitfig} shows an example of numerical results at a general point
of the phase diagram. The following formula gives an excellent fit to the
numerical data throughout the phase diagram (except very close to the lines
with the $1/\sqrt{r}$ behavior):
 \be
 \rho^x(r) = \left\{ \begin{array}{ll}
 (a_1\cos(k_h r)+a_2\cos(k_\lambda r))/r & \mbox{for r even,} \cr
 (a_3\cos(k_h r)+a_2\cos(k_\lambda r))/r~~& \mbox{for r odd.} \cr
 \end{array} \right.
 \label{bela}
 \ee
The $a_i$ coefficients are functions of $h$ and $j^E$,
and for the \one phase $a_1=a_2$ seems to be valid. 

As one can see from (\ref{bela}), the amplitude of the $1/r$
decay is modulated with the critical wavenumbers $k_h$ and $k_\lambda$. 
On the `high-symmetry' transition line we have $k_h=k_\lambda$ and the 
transition across this line takes us from region \one where 
$k_h<k_\lambda$ to region \two where $k_h>k_\lambda$. Thus one can view
the `high-symmetry' line as a line of degeneracy 
where two characteristic wavelengths of the system become equal. 

\begin{figure}[htb]
\centerline{
        \epsfxsize=8.0cm
        \epsfbox{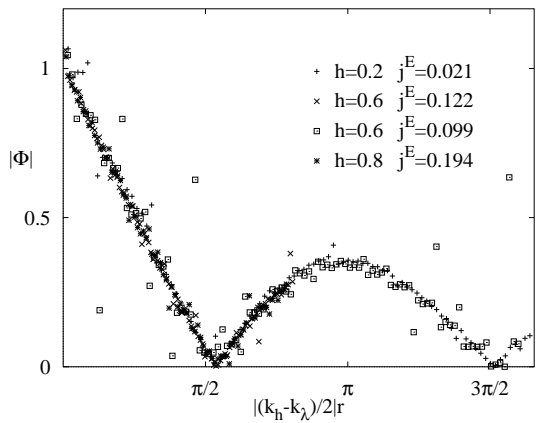}
           }
	\vspace{0.5cm}
\caption{ The absolute value of the scaled $\rho^x$ correlation function at
        four points of the phase diagram near the boundary between 
        regions \one and \two. Two points are above and two are
	below this `high'-symmetry line. 
        The data points showing large deviations from scaling come
	from arguments where both the numerator and the denominator in
	(\ref{scale1}) are close to zero.
}
\label{scale1fig}
\end{figure}

This transition may resemble transitions arising from competing wavelengths 
but, actually, here we do not have a competition between
$k_h$ and $k_\lambda$. Due to the product form of $\Lambda_k$, 
$k_h$ is independent of $\lambda$ while
$k_\lambda$ is independent of $h$. Nevertheless, this transition does 
have similarities with second order transitions in that the correlations 
decay more slowly ($1/r \rightarrow 1/\sqrt{r}$) and, furthermore,
one can observe scaling upon approach of the transition line. In order to see
this, let us assume that the distance from the transition line, 
$k_h-k_\lambda$, provides the single diverging lengthscale which 
generates the $1/r \rightarrow 1/\sqrt{r}$ crossover in correlations.
Then one should observe scaling when plotting the following ratio:
 \be
 \frac{\rho^x(r;h,j^E)}{\rho^x(r; h_c, j^E_c)} = 
  \Phi \left( {|k_h-k_\lambda| \over 2}r\right)~,
 \label{scale1}
 \ee
where $(h_c,j^E_c)$ is a point on the transition line  and the 
$(h\rightarrow h_c, j^E\rightarrow j^E_c)$ limit is taken.
Note that the long range behavior of the denominator ($\rho^x$ 
on the phase transition line) is known (\ref{bound}). 
As one can see from Fig.\ref{scale1fig}, the data collapse is excellent thus
supporting the assumption of scaling (\ref{scale1}). 

It is interesting to note that the scaling function appears 
to be the same on the both sides of the phase-transition line.
Furthermore, $\Phi$ is independent of the crossing point $(h_c,j^E_c)$  
unless we are close to the zero-field equilibrium point $(h=0,j^E=0)$. 
In this sense we have the kind of universality usually observed in 
critical phase transitions.

The equilibrium point $(h=0,j^E=0)$ is the endpoint of the 
`high'-symmetry line. At this point, the ground state symmetry is higher
than on the line and so we may expect that, provided scaling was still
present, the scaling function would be different. This is indeed what we 
observe.
\begin{figure}[htb]
\centerline{
        \epsfxsize=8.0cm
        \epsfbox{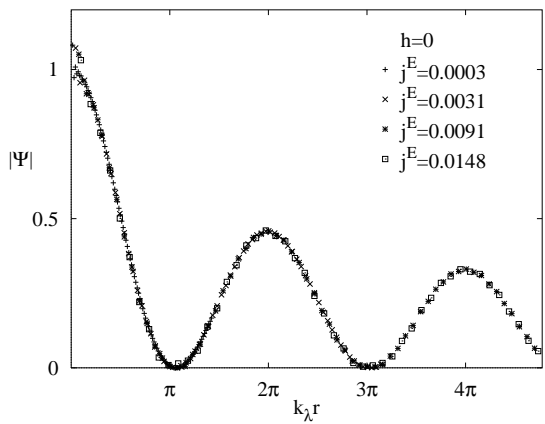}
           }
\vspace{0.3cm}
\caption{ The absolute value of the scaled $\rho^x$ correlation functions on
        the $h=0$ line near $j^E=0$. }
\label{scale4fig}
\end{figure}
Approaching the point $(h=0,j^E=0)$ along 
the $(h=0,j^E\rightarrow 0)$ line, 
one has again a diverging lengthscale proportional to $1/k_\lambda$ and
one can search again for scaling in the correlation function 
 \be
 {\rho^x(r;0, k_\lambda) \over \rho^x(r;0, 0)}=\Psi(k_\lambda r)~.
 \label{scale4}
 \ee
As shown in Fig.\ref{scale4fig}, scaling is indeed seen and the scaling
function is significantly different from that 
found away from the $(h=0,j^E=0)$ point.

The numerical results presented above (as well as  
other data gathered in our explorations of the phase diagram) 
suggest strongly that 
$\rho^x \propto 1/r$ for generic current-carrying states. Slower decay,
$\rho^x \propto 1/\sqrt{r}$ is observed only on the boundaries 
of region \two and the crossover between the $1/r$ and $1/\sqrt{r}$ 
behaviors can be understood in terms of single-lengthscale scaling.
It is intriguing that there is a simple correspondence between 
the types of decay of correlations and the "band
structure" of the ground state. The lines of slower
decay of correlations coincide with those lines where the 
ground state is build by a
single band of excitations in momentum space, whereas in all regions of
$1/r$-decay, the filling pattern of the ground state splits into two separate
bands (Fig. 1).

Regarding the
interplay of currents and correlations these results leave us
with the following conclusions. First, we find that the 
large-distance correlations are not necessarily increased by
switching on a current. Second, it is found 
that the equilibrium power-law correlations are not destroyed by the current, 
only the exponent in the power law is increased. 
This strengthens previous observations 
that currents and power-law correlations are intimately related.
Third, we find that the increase of current may lead to interesting
phase-transition like behavior related to the {\em increase of symmetry}
at special values of the current. 


\section{Final remarks}
\label{sec:final}

A general conclusion we can draw from the present 
study of the transverse $XY$ model and from comparison with 
the results on the transverse Ising model \cite{trising} 
is that currents appear to generate and maintain power-law correlations. 
An interesting feature of $XY$ model which may also have some generality 
is the increase of the 
symmetry of the ground state at special values of the energy current. 
This feature should certainly be searched for in other models as well
as in experiments. 
It should be recognized, however, that both the $XY$ and the transverse Ising
models are integrable and, consequently, they  
are special 
in that conductivity and, in particular, the thermal conductivity is ideal 
for them (not only at zero but also at nonzero temperatures)
\cite{{Shastry},{Zotos},{Saito}}. Thus it is an important next step 
to find out whether nonintegrable models have the same connection between 
currents and power-law correlations and, furthermore, whether they show 
any additional general features.

\section{Acknowledgements}

We thank J. Cardy, F. Essler, and L. Sasv\'ari for helpful
discussions. Z.R. thanks for partial support by 
the Hungarian Academy of Sciences (Grant OTKA T 019451) and by
the EPSRC, United Kingdom (Grant L58088).


\end{multicols}

\begin{references}
\bibitem{magcurr} Note that although only magnetization 
                 current flows in these states, they nevertheless
                 differ from the current-carrying states studied 
                 in Section \ref{sec:mag}. There the expectation 
                 value of the interaction 
                 part of the current takes a nonzero value. The only 
                 current-carrying state identical in both models
                 is at the parameter values $h=1$ and $j^M=j^E=1/\pi$.
\end{references}
\end{document}